# Closed-Form PI and PID Tuning of All-Pole Plants up to Third Order for Monotonic, Minimum-Settling Step Responses

Senol Gulgonul
*Department of Electrical and Electronics Engineering, Ostim Technical University, Ankara, Turkey*
senol.gulgonulostimteknik.edu.tr

## Abstract

A unified, closed-form analytical PI/PID tuning method is presented for all-pole plants up to third order that yields a strictly monotonic (zero-overshoot) step response with minimum settling time. The design target is the binomial closed loop $p^n/(s+p)^n$, which is monotonic with robustness depending only on the order n. Because a fixed PI/PID cannot assign the closed-loop poles and the controller zeros independently, realizing this target exactly requires the controller zeros to be cancelled, which forces the controller numerator to divide the plant denominator. Carrying this principle through shows that an exact, real-gained solution exists for any stable plant precisely up to second order with a PI controller and third order with a PID controller; the residual binomial factor acquires a complex pair (damping √3/2) beyond that, which a generic plant does not contain. Explicit gains are derived for first-order plants (PI), second-order plants with real and complex poles (PI and PID), and third-order plants with three real poles and with one real pole plus a complex pair (PID). The second-order PI case is treated in full as the lowest-order instance. The freedom of the coincident designs is shown to be bounded: a quadratic nonnegativity condition gives the exact window of the design pole for strict monotonicity, which collapses at the pole-ratio-2 changeover for real poles and is nonempty for damping ratios above approximately 0.443 for complex poles. Monotonicity guarantees $M_t = 1$, hence $M_s \leq 2$, phase margin ≥ 60°, and gain margin ≥ 6 dB, tightening to universal constants for the binomial family. Load-disturbance attenuation obeys IAEd = 1/Ki, making the cost of cancellation explicit, and comparisons with SIMC, the CHR zero-overshoot rule, and deadbeat-fitted explicit formulas quantify the trade: at matched maximum sensitivity the proposed design settles faster than SIMC on the third-order example, and the monotonicity guarantee is obtained with markedly lower controller gains and peak control effort. Numerical verification confirms all designs.

**Keywords:** PI controller; PID controller; pole placement; pole–zero cancellation; monotonic response; zero overshoot; settling time; maximum sensitivity; robustness; load disturbance; SIMC.

## 1. Introduction

The proportional–integral–derivative (PID) controller remains the dominant feedback controller in industry. Systematic tuning began with Ziegler and Nichols [1] and was refined by many authors [2], [3]. Model-based analytical rules such as Internal Model Control and its SIMC form derive the gains directly from the plant and a single closed-loop time constant [4], while Åström and Hägglund established the maximum-sensitivity $M_s$ as the standard robustness measure and gave pole-placement formulas for PI control of low-order plants [5]. The magnitude- and symmetric-optimum school shapes the closed-loop frequency response toward unity over a band [6–8], and margin-specification methods assign gain and phase margins directly [9–11], with the complete stabilizing PID set characterized in [12]; in all of these the time-domain shape of the step response remains implicit. Optimal and explicit gain formulas have also been fitted to integral-error criteria [13] and to second-order plants [14], and the Ziegler–Nichols step-response route was modernized as AMIGO [15]. The present work inverts the usual order: strict monotonicity of the time-domain response is the specification, and the robustness figures follow from it as universal constants.

A frequent requirement in precision positioning, thermal regulation, and dosing is that the step response be non-overshooting and monotonically non-decreasing, since overshoot can cause damage or quality loss, while settling time must be small for throughput. These objectives compete. Kim, Keel and Bhattacharyya controlled overshoot and speed of all-pole systems directly through the characteristic ratios and a generalized time constant, proving that repeated-pole configurations are overshoot-free [16]; their procedure, however, employs a state-feedback plus feedforward or two-parameter output structure rather than a fixed PI/PID, and does not provide explicit gains for a standard PID under a minimum-settling criterion. Many higher-order processes are routinely reduced to first-, second- or third-order models for design [17]. Nguyen and Nguyen assigned overshoot and settling time with a PID by cancelling both plant poles of a real-pole second-order plant, reducing the loop to first order with the proportional gain setting the speed [18]; complex poles, robustness guarantees, and orders beyond the second are not addressed there; Albatran et al. gave explicit but numerically fitted PI/PID formulas for underdamped second-order plants without a strict monotonicity guarantee [19].

The present work treats the second-order PI case in full as the lowest-order instance and extends the construction to PID control and to third-order plants. For all-pole plants up to third order a closed-form PI/PID tuning is derived that is strictly monotonic and minimum-settling, covers real and complex poles, exposes exactly when a free design parameter exists, and inherits guaranteed robustness from monotonicity itself. The third order is shown to be the precise boundary beyond which a standard PID can no longer realize the cancellation that minimum settling demands. The practical motivation is direct: in thermal processing, batch operation, and motion systems with hard position constraints, overshoot is unacceptable and commissioning time is scarce; one-line gain formulas with a guaranteed monotonic response, fixed robustness, and unit-order gains address exactly this setting; tuning aimed jointly at settling time and overshoot is likewise the stated goal in high-performance drive control [20]. The specific contributions are: the exact characterization of when the binomial target is realizable: up to second order with PI and third order with PID, set by the residual factor g(s) being real-rooted only for $n \leq 2$; the feasibility condition $\zeta > \sqrt{3}/2$ for fixed-σ PI control of complex second-order plants, together with the exact monotonicity window of the free-σ PID branch (nonempty for $\zeta \gtrapprox 0.443$); the piecewise cancellation/coincident rule with its changeover at $T_1/T_2 = 2$ and quantified settling-time gains; and the closed-form universal robustness constants of the binomial family, including $GM(n) = 1 + \sec^n(\pi/n)$.

## 2. Problem Statement and the Binomial Target

Consider a stable all-pole plant and a PID controller in unity feedback,

$$P(s) = \frac{1}{M(s)}, \quad C(s) = K_p + \frac{K_i}{s} + K_d\, s = \frac{N(s)}{s}, N(s) = K_d s^2 + K_p s + K_i \tag{1}$$

where M(s) is a monic polynomial of degree m with poles in the open left half-plane and N(s) is the controller numerator (one zero for PI, two for PID). The closed-loop transfer function and its characteristic polynomial are

$$T(s) = \frac{N(s)}{s\,M(s) + N(s)}, \quad Q(s) = s\,M(s) + N(s), \quad deg\, Q = m + 1, \tag{2}$$

so the controller numerator N(s) also appears as the numerator of T, placing its zeros in the closed loop. The step response y(t) is monotonic when

$$dy/dt = h(t) \geq 0, \quad for\ all\ t \geq 0, \tag{3}$$

where h(t) is the impulse response.

### *2.1 The binomial target and the cancellation principle*

A closed-loop shape that is monotonic and whose robustness is independent of its speed is the binomial form

$$T^*(s) = \frac{p^n}{(s+p)^n} \tag{4}$$

whose impulse response $t^{n-1}e^{-pt}$ is non-negative for every n, and whose maximum sensitivity, complementary sensitivity and stability margins depend only on n, not on p. A shifted form $(s+z_1)(s+z_2)p^n/z_1z_2(s+p)^n$ with the zeros of PI-PID controller to the left of −p remains monotonic; indeed, for a single zero at −z the step response becomes $y(t) + h(t)/z \geq y(t)$, so for a fixed pole pattern a retained left-half-plane zero settles no later. The zeros, however, are not free: with a fixed PI/PID the closed-loop poles and the controller zeros cannot be assigned independently, and retaining the zeros constrains how deep the closed-loop poles can be placed. The pure binomial target is realized exactly when the controller zeros are cancelled, i.e. $T(s) = T^*(s)$. Equating (2) to (4) and clearing denominators gives

$$N(s)[(s+p)^n - p^n] = p^n s\, M(s), \qquad (s+p)^n - p^n = s \cdot g(s)$$

$$g(s) = \frac{[(s+p)^n - p^n]}{s}, \qquad N(s) \cdot g(s) = p^n\, M(s)$$

$$M(s) = N(s) \cdot g(s)/p^n$$

$$g(s) = 1\ (n = 1), g(s) = s + 2p\ (n = 2), g(s) = s^2 + 3ps + 3p^2\ (n = 3) \tag{5}$$

Thus N(s) must divide M(s): the controller zeros sit on plant poles and they are cancelled for pure binomial target transfer function. Degree matching fixes $n = m - q + 1$, where $q = \deg N(s)$, giving $n = m$ for PI and $n = m - 1$ for PID. A real-gained solution exists only when g(s) is available as a real factor of the plant. But g(s) is real-rooted only for $n \leq 2$.

Hence the cancellation construction succeeds for an arbitrary stable plant exactly up to second order with PI ($n \leq 2$) and third order with PID ($n = m - 1 \leq 2$). This single fact is the origin of both the third-order boundary and the $\zeta > \sqrt{3}/2$ limit for complex second-order plants.

### *2.2 Coincident placement for clustered poles*

When the plant poles are clustered, a faster monotonic response is obtained by instead placing all $m + 1$ closed-loop poles at one location $-\sigma$, keeping the controller zeros (which then lie to the left of $-\sigma$). Matching the $s^m$ coefficient of $Q(s) = (s + \sigma)^{m+1}$, which the controller cannot alter, fixes

$$\sigma = c_{m-1} / (m + 1) = \frac{\Sigma\ (reciprocal\ time\ constants)}{m + 1}, \tag{6}$$

where $c_{m-1}$ is the coefficient of $s^{m-1}$ in M(s); each case below specializes (6) accordingly. This branch requires the controller numerator degree to be $m - 1$, again limiting the exact construction to $m \leq 3$. Both branches are monotonic; the faster one is used (cancellation for separated poles, coincident placement for clustered poles), and they coincide at the changeover.

## 3. First- and Second-Order Plants

### *3.1 First-order plant (PI)*

For first order plant P(s) a PI controller whose zero cancels the plant pole ($n = 1$) gives, after cancellation, the first-order closed loop T(s), monotonic for any speed [5]:

$$P(s) = \frac{1}{sT_1 + 1}, T(s) = \frac{1}{\lambda s + 1}, \qquad K_p = \frac{T_1}{\lambda}, \quad K_i = \frac{1}{\lambda}, \quad K_d = 0, \tag{7}$$

where $\lambda$ is the closed-loop time constant and is freely chosen.

### *3.2 Second-order plant, real poles*

A second-order real plant P(s) with $T_1 \geq T_2$ admits two monotonic designs, and the pole ratio $T_1/T_2$ decides which settles faster: pole–zero cancellation of the slow pole, which sets a critically damped double pole, and coincident placement of all three closed-loop poles at a common $-\sigma$ while the controller zero is retained at $-z$. When the poles are well separated ($T_1/T_2 \geq 2$) the cancellation branch applies: the PI zero cancels the slow pole and the gain sets the double pole at $-1/(2T_2)$, giving the pure binomial T(s) ($n = 2$):

$$P(s) = \frac{1}{(sT_1 + 1)(sT_2 + 1)}, \; T(s) = \frac{1}{(2T_2 s + 1)^2}, \qquad K_p = \frac{T_1}{4\, T_2}, \quad K_i = \frac{1}{4\, T_2} \tag{8}$$

When the poles are clustered ($T_1/T_2 < 2$), coincident placement of a triple pole is faster, and the location of the pole is forced rather than chosen. With a PI the closed-loop denominator is $T_1T_2s^3 + (T_1 + T_2)s^2 + (1 + K_p)s + K_i$, and no controller gain reaches the $s^2$ coefficient; matching it to $T_1T_2(s + \sigma)^3$ therefore fixes $3T_1T_2\sigma = T_1 + T_2$, the case m = 2 of (6), and the two remaining coefficients yield the gains:

$$\sigma = (T_1 + T_2)/(3T_1T_2), \qquad T(s) = (Kps + Ki)/T_1T_2(s + \sigma)^3$$

$$K_p = \frac{T_1^2 - T_1T_2 + T_2^2}{3\,T_1T_2}, \qquad K_i = \frac{(T_1 + T_2)^3}{27\,T_1^2T_2^2}. \tag{9}$$

The controller zero of this design, $z = K_i/K_p$, lies at

$$z = (T_1 + T_2)^3 / [9\,T_1T_2(T_1^2 - T_1T_2 + T_2^2)]\,, \tag{10}$$

The coincident closed loop $T(s) = K\,K_p(s + z)/(s + \sigma)^3$ is monotonic only when the zero lies to the left of the triple pole, that is $z \geq \sigma$; a zero nearer the origin adds a term that rises before decaying and produces overshoot. Imposing $z \geq \sigma$ and cancelling the positive factor $3\,T_1T_2/(T_1 + T_2)$ gives

$$(T_1 + T_2)^2 \geq 3(T_1^2 - T_1T_2 + T_2^2)\,, \quad \text{i.e.} \quad 2T_1^2 - 5T_1T_2 + 2T_2^2 \leq 0\,, \tag{11}$$

Dividing by $T_2^2$ and writing $r = T_1/T_2 \geq 1$ reduces this to $2r^2 - 5r + 2 \leq 0$, whose roots are $r = 1/2$ and $r = 2$; with $r \geq 1$ the feasible range is $1 \leq T_1/T_2 \leq 2$. The coincident design is therefore monotonic exactly for clustered poles, $T_1/T_2 \leq 2$, while pole-zero cancellation is used for separated poles, $T_1/T_2 \geq 2$. Throughout $1 \leq T_1/T_2 < 2$ both designs are monotonic, but the coincident pole $-\sigma$ is deeper than the cancellation pole $-1/(2T_2)$, so the coincident response settles faster and is preferred; the dominant-pole ratio $3r/[2(1 + r)]$ bounds the gain, and exact simulation gives a 2% settling-time reduction of 17.1% at $T_1/T_2 = 1$, 10.1% at 1.2, and 3.6% at 1.5, decaying to zero as $T_1/T_2 \to 2$. At the boundary $T_1/T_2 = 2$ the zero meets the triple pole, $z = \sigma$, the two designs give identical gains, and the closed loop reduces to the critically damped double pole; the piecewise tuning is thus continuous at the changeover. The same rule, that the controller zero must remain to the left of the repeated closed-loop pole, fixes the changeover at every order, with a threshold set by the plant.

Adding derivative action (a full PID) releases $\sigma$ as a free parameter in the same T(s) recovering the PI triple-pole solution at $K_d = 0$:

$$T(s) = N(s)/T_1T_2(s + \sigma)^3$$

$$K_d = 3\,T_1T_2\,\sigma - (T_1 + T_2)\,, \quad K_p = 3\,T_1T_2\sigma^2 - 1\,, \quad K_i = T_1T_2\sigma^3\,. \tag{12}$$

The release of $\sigma$ is bounded. The impulse response of (12) factors as $h(t) = e^{-\sigma t}\,q(t)/(T_1T_2)$, with the quadratic

$$q(t) = K_d + (K_p - 2\sigma K_d)\,t + \tfrac{1}{2}\,\sigma M(-\sigma)\,t^2\,, \tag{13}$$

Strict monotonicity is equivalent to $q(t) \geq 0$ for all $t \geq 0$. Since $M(-\sigma) = (T_1\sigma - 1)(T_2\sigma - 1)$ and $K_p - 2\sigma K_d \geq 0$ throughout the interval, this holds exactly for $\sigma_0 \leq \sigma \leq 1/T_1$: the coincident pole may be deepened only up to the slow plant pole, and the window collapses to a point at the changeover ratio $T_1/T_2 = 2$. For faster monotonic responses the cancellation branch below applies.

Alternatively, a PID can cancel both plant poles and reduce the loop to the single pole $T(s) = 1/(\lambda s + 1)$, the minimum-settling design, with $K_d = T_1T_2/\lambda$, $K_p = (T_1 + T_2)/\lambda$, $K_i = 1/\lambda$, and $\lambda$ free; this is the second-order PID entry of Tables 2–4, and it coincides with the zero-overshoot design of Nguyen and Nguyen [18], which treats real-pole plants only.

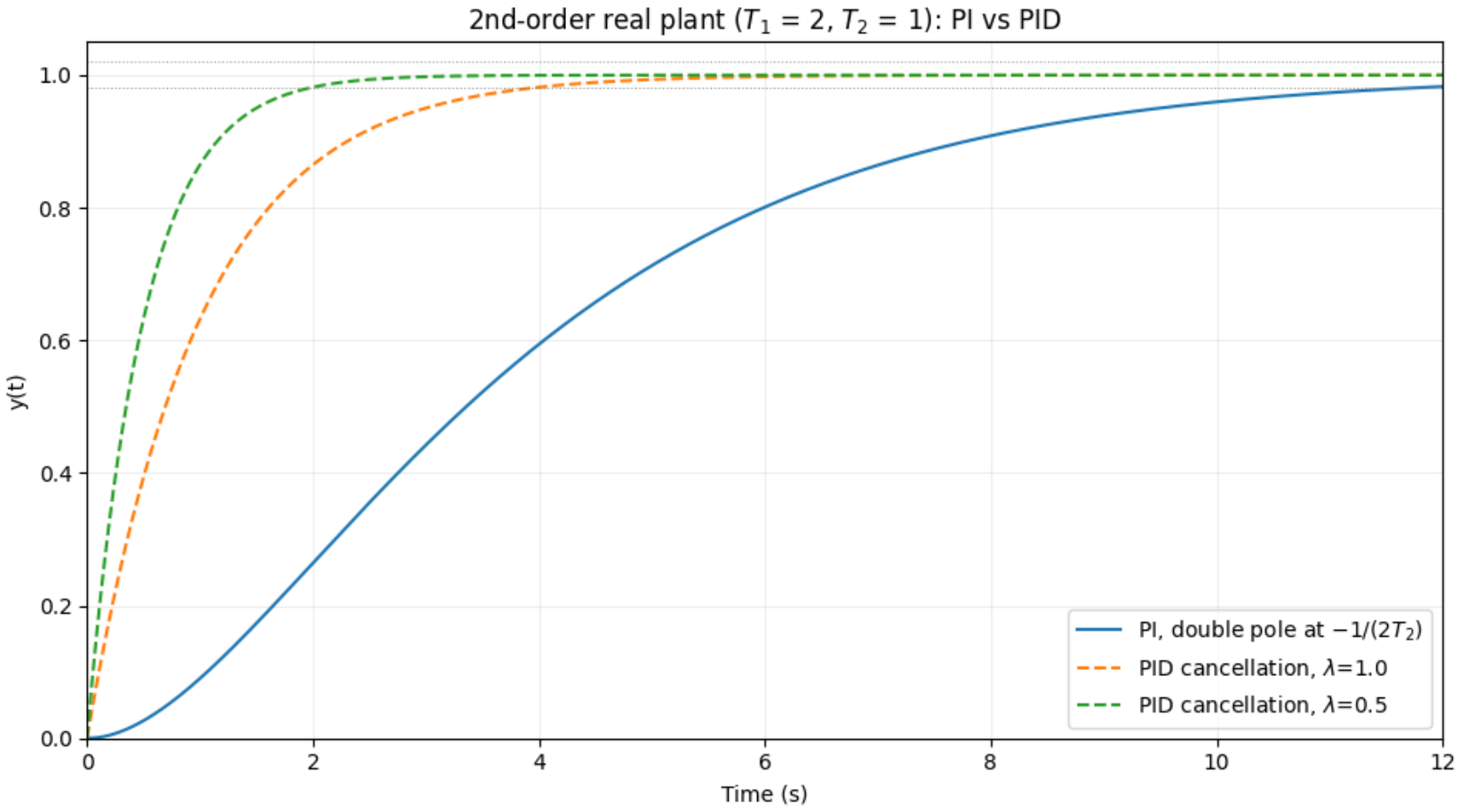

*Fig. 1. Second-order real plant: the PI solution (forced σ) and faster monotonic PID designs (pole-zero cancellation, free λ).*

### 3.3 Second-order plant, complex poles

For P(s) second order plants with $\zeta < 1$, coincident placement gives:

$$P(s) = 1/(T^2s^2 + 2\zeta Ts + 1),\ \sigma = 2\zeta/(3T),\ T(s) = (K_ps + K_i)/T^2(s+\sigma)^3$$

$$K_p = \frac{4\zeta^2 - 3}{3}, \quad K_i = \frac{8\zeta^3}{27\,T}, \quad K_d = 0, \tag{14}$$

which is feasible (positive gains, monotonic) if and only if

$$K_p > 0 \Rightarrow \zeta > \frac{\sqrt{3}}{2} \approx 0.866, \tag{15}$$

the same $\sqrt{3}/2$ that arises as the damping of the residual factor g in (5). For $\zeta \le \sqrt{3}/2$, a PID with free σ restores feasibility over a bounded window of σ:

$$T(s) = N(s)/T^2(s+\sigma)^3$$

$$K_d = 3T^2\sigma - 2\zeta T, \quad K_p = 3T^2\sigma^2 - 1, \quad K_i = T^2\sigma^3. \tag{16}$$

Here, too, σ is free only within a window. With $h(t) = e^{-\sigma t} q(t)/T^2$ and q(t) as in (13), where now $M(-\sigma) = T^2\sigma^2 - 2\zeta T\sigma + 1 > 0$, strict monotonicity holds if and only if $K_d \ge 0$ and $K_p - 2\sigma K_d \ge -[2\sigma M(-\sigma)K_d]^{1/2}$. For $\zeta = 0.6$ (T = 1) the window is $0.567 \le \sigma \le 1.151$, and the window is nonempty exactly for $\zeta \ge \zeta^* \approx 0.443$, the smallest damping at which the quadratic-nonnegativity condition admits a solution, obtained numerically. Evaluating the linear coefficient at the PI point $\sigma = 2\zeta/(3T)$ gives $K_p - 2\sigma K_d = 4\zeta^2/3 - 1$, recovering (15). Below $\zeta^*$ no member of this family is strictly monotonic, although non-overshooting responses persist; the strict specification then has no solution within the coincident structure.

A pole–zero cancellation PID also exists for the complex pair: choosing the controller numerator as $(T^2s^2 + 2\zeta Ts + 1)/\lambda$, i.e. $K_d = T^2/\lambda$, $K_p = 2\zeta T/\lambda$, $K_i = 1/\lambda$, gives $T(s) = 1/(\lambda s + 1)$ with all gains positive for any $0 < \zeta < 1$. It is nevertheless excluded from Table 2 for two reasons. First, the cancelled underdamped pair remains in the load-disturbance path: at matched settling time (λ = 1.33 against the window-limit design σ = 1.151 for ζ = 0.6) the disturbance response rings (it changes sign, so $IAE_d$ exceeds $1/K_i$), with peak deviation 0.44 against 0.20 and $IAE_d$ = 1.38 against 0.66 for the coincident design. Second, strict monotonicity of the setpoint response then rests on exact cancellation of an oscillatory pair; under parameter mismatch the property is lost, mildly at moderate damping and increasingly as ζ decreases, whereas the coincident design retains it with margin. For real poles the same cancellation is benign and is used throughout; for complex pairs the coincident window is the recommended route.

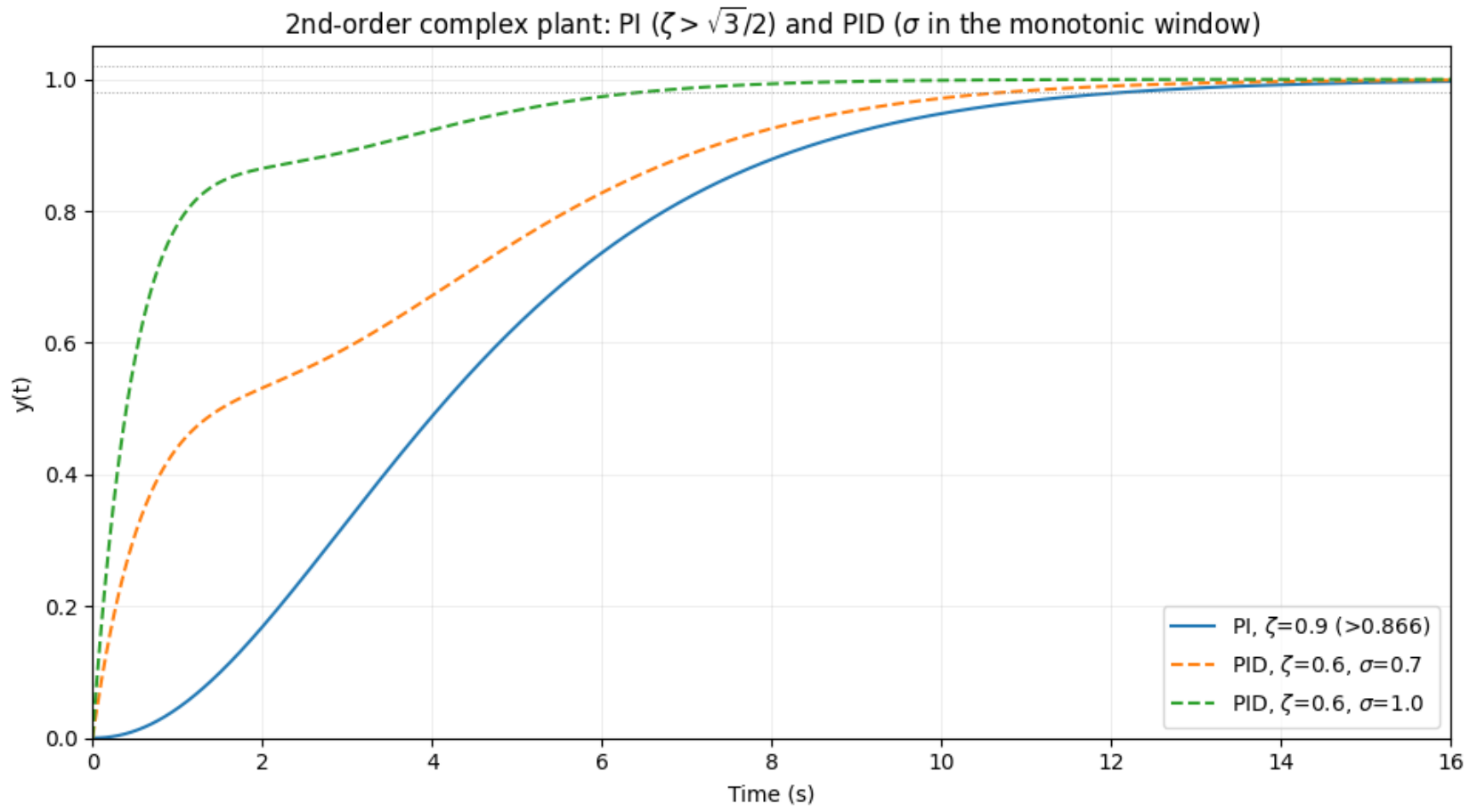

*Fig. 2. Second-order complex plant: PI for ζ > √3/2, and PID with σ inside the monotonicity window (ζ = 0.6).*

# 4. Third-Order Plants

## *4.1 Three real poles*

For P(s) third order plants the PID is the minimal controller (n = m − 1 = 2, so g = s + 2p is still real). Coincident placement of a quadruple pole gives, from (6),

$$P(s) = 1/(sT_1 + 1)(sT_2 + 1)(sT_3 + 1)$$

$$\sigma = \frac{T_1T_2 + T_1T_3 + T_2T_3}{4\,T_1T_2T_3}, \tag{17}$$

and T(s) with

$$T(s) = N(s)/T_1T_2T_3(s + \sigma)^4$$

$$K_d = 6T_1T_2T_3\sigma^2 - (T_1 + T_2 + T_3)\,,\ K_p = 4T_1T_2T_3\sigma^3 - 1\,,\ K_i = T_1T_2T_3\sigma^4\,. \tag{18}$$

For well-separated poles the faster, minimum-settling choice cancels the two slower poles and keeps the fastest, placing the binomial double pole at $-1/(2T_3)$, where $T_3$ is the smallest time constant. Writing the two cancelled constants as $T_1$ and $T_2$, $K_d = T_1T_2/(4T_3)$, $K_p = (T_1 + T_2)/(4T_3)$, $K_i = 1/(4T_3)$, and no free parameter remains; coincident placement (15) is then reserved for clustered poles, exactly as in the second-order case.

## *4.2 One real pole and a complex pair*

For P(s) having one real pole and a complex pair the same construction gives:

$$P(s) = \frac{1}{(T_3s + 1)(T^2s^2 + 2\zeta Ts + 1)},\ \ \sigma = \frac{T + 2\zeta T_3}{4TT_3}, T(s) = N(s)/T_3T^2(s + \sigma)4$$

$$K_d = 6T_3T^2\sigma^2 - (T_3 + 2\zeta T)\,,\ K_p = 4T_3T^2\sigma^3 - 1\,,\ K_i = T_3T^2\sigma^4\,. \tag{19}$$

The design is feasible while the complex pair is not too lightly damped; a strongly underdamped pair drives the gains negative, and since the PID is already minimal for the third order no free parameter remains to recover it.

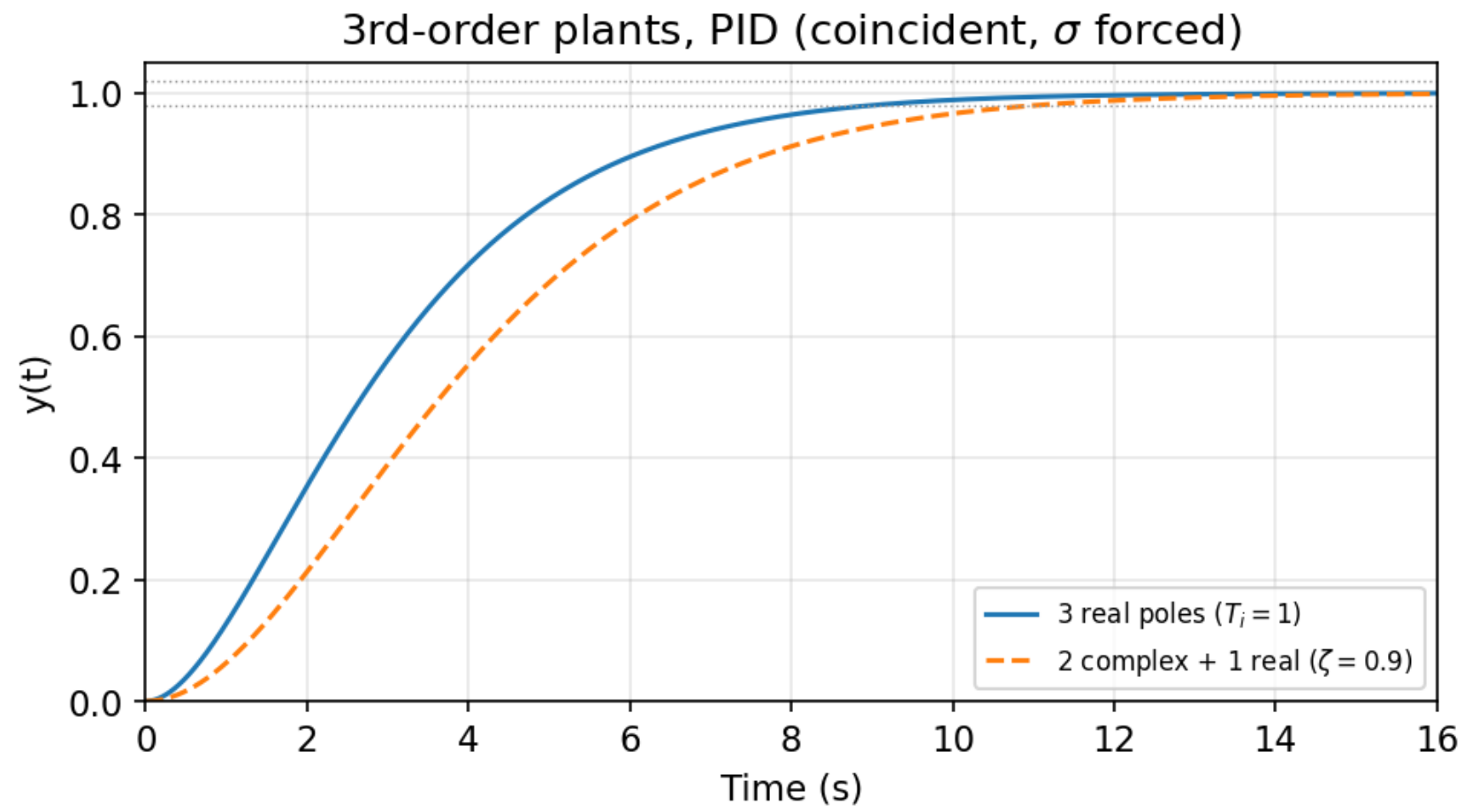

*Fig. 3. Third-order plants: monotonic PID responses for three real poles and for a complex pair plus a real pole.*

## 5. Robustness and Settling Time

Every design above is monotonic. A non-negative impulse response with unit final value bounds the complementary sensitivity, since

$$|T(j\omega)| \;=\; |\int \, h(t)\, e^{-j\omega t}\, dt| \;\leq\; \int \, h(t)\, dt \;=\; T(0) \;=\; 1, \tag{20}$$

so $M_t = \max|T(j\omega)| = 1$. The standard peak-to-margin inequalities then give, with no further design effort,

$$M_s \;\leq\; 2\,, \qquad PM \;\geq\; 60°\,, \qquad GM \;\geq\; 6\,dB. \tag{21}$$

For the binomial family $T = p^n/(s + p)^n$ these tighten to universal constants of n alone, which follow from the normalized loop. In the scaled frequency $u = \omega/p$ the loop transfer function is

$$L(ju) = 1 \,/\, [(1 + ju)^n - 1] \tag{22}$$

which depends only on n, so every robustness measure is a universal function of n. The sensitivity $S = 1 - T$ has squared magnitude

$$|S(ju)|^2 = |(1 + ju)^n - 1|^2 \,/\, (1 + u^2)^n \tag{23}$$

whose maximum is the maximum sensitivity $Ms = 2/\sqrt{3} \approx 1.155$ for $n = 2$ (attained at $u = 1/\sqrt{2}$), and $9/7 \approx 1.286$ and $\approx 1.381$ for $n = 3$ and 4. For $n = 2$ the gain crossover $16u^4 + 16u^2 - 1 = 0$ has positive root 0.243, giving $PM = 90° - \arctan(0.243) = 76.35°$. For general n the crossover $|L(ju)| = 1$, with $\theta = n{\cdot}\arctan u$, reduces to

$$\sec^n(\theta/n) = 2\cos\theta\,, \qquad PM(n) = 180° - 2\theta^* \tag{24}$$

whose root $\theta^*$ also gives the crossover frequency $\omega c/p = \tan(\theta^*/n)$, so $PM = 71.25°$ and $68.58°$ (with $\omega c/p = 0.327$ and $0.248$) for $n = 3$ and 4. The phase reaches $-180°$ at $u = \tan(\pi/n)$, where $|1 + ju|^n = \sec^n(\pi/n)$, so the gain margin is

$$GM(n) \;=\; 1 \;+\; sec^n(\pi/n)\,, \qquad n \;\geq\; 2, \tag{25}$$

which equals $\infty$ for $n \leq 2$, 9 (19.08 dB) for $n = 3$ and 5 (13.98 dB) for $n = 4$, descending to a hard floor of 2 (6.02 dB) as $n \to \infty$ since $\sec^n(\pi/n) \to 1$. Finally, the binomial step error depends only on n, so its 2% level fixes a universal constant $c_n$, and the settling time scales as

$$T_s = c_n \,/\, \sigma\,, \qquad c_2 = 5.83\,,\; c_3 = 7.52\,,\; c_4 = 9.08 \tag{26}$$

Table 1 lists the universal measures for the orders arising up to third-order plants.

*Table 1. Universal robustness of the binomial family T = p^n/(s + p)^n.*

| n | Mt | Ms | PM (°) | GM |
|---|---|---|---|---|
| 1 | 1.000 | 1.000 | 90.0 | ∞ |
| 2 | 1.000 | 1.155 | 76.35 | ∞ |
| 3 | 1.000 | 1.286 | 71.25 | 19.1 dB |
| 4 | 1.000 | 1.381 | 68.58 | 14.0 dB |

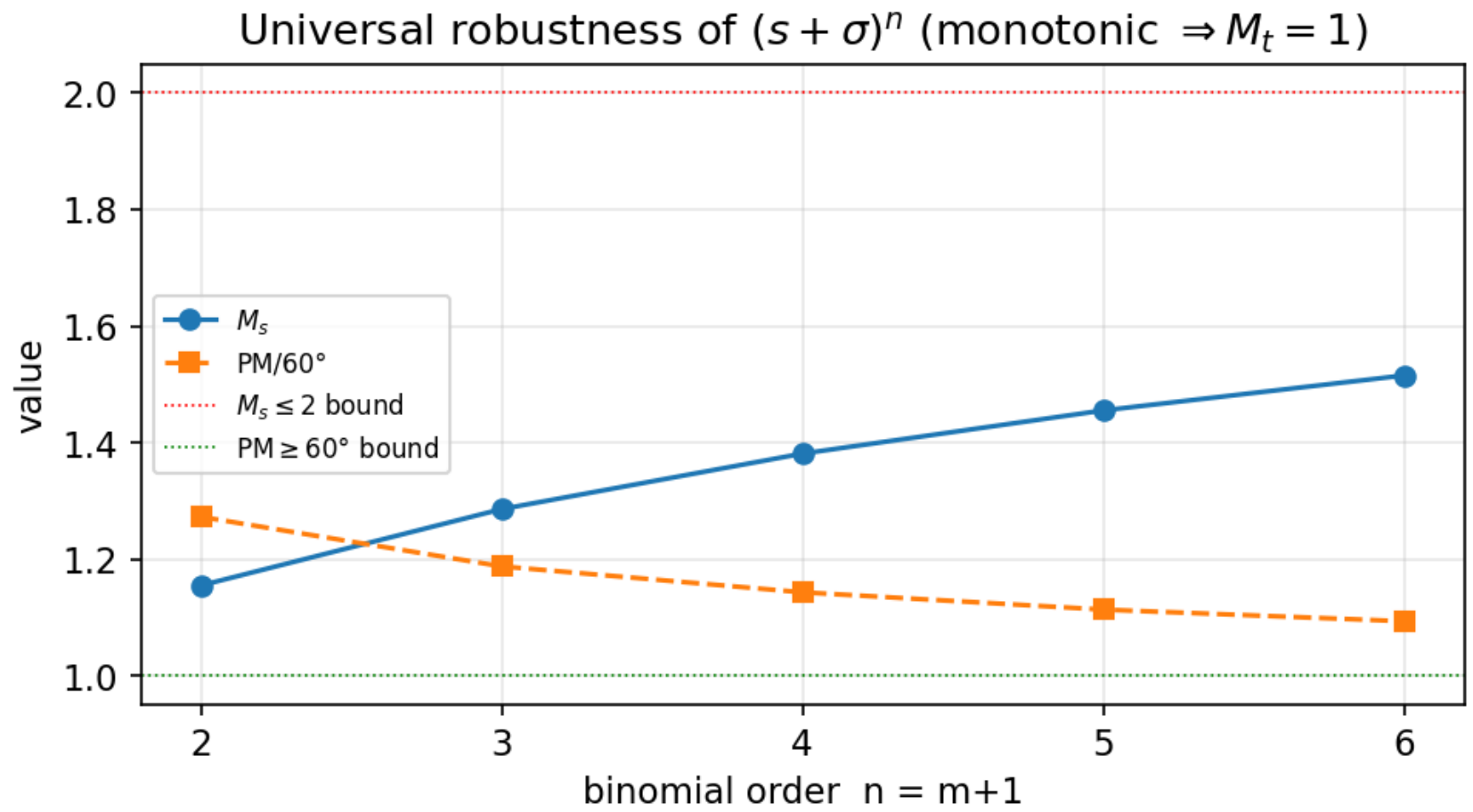

*Fig. 4. Universal robustness of the binomial family: Ms and PM versus order n, with the monotonicity floors.*

## 6. Numerical Verification

Table 2 summarizes the cases, Table 3 collects the closed-form gains, and Table 4 lists representative designs and the resulting closed-loop measures, computed from the closed-form expressions and confirmed by simulation. All responses are strictly monotonic with $M_t = 1$.

*Table 2. Summary of cases for all-pole plants up to third order.*

| Plant order | Poles | Controller | Method | Pole location |
|---|---|---|---|---|
| 1 | real | PI | cancellation | free |
| 2 | real | PI | cancellation | fixed |
| 2 | real | PID | cancellation | free |
| 2 | complex | PI ($\zeta > \sqrt{3}/2$) | coincident | fixed |
| 2 | complex | PID | coincident | free (window) |
| 3 | 3 real | PID | cancellation | fixed |
| 3 | 2 complex + 1 real | PID | coincident | fixed |

*Table 3. Closed-form gains of the piecewise design ($T_1 \geq T_2 \geq T_3$; $\lambda > 0$ is the closed-loop time constant of a cancellation design; $\sigma$ is the coincident pole with $\sigma_0 = (T_1+T_2)/(3T_1T_2)$; all responses are strictly monotonic).*

| Plant | C | Branch | $K_p$ | $K_i$ | $K_d$ | Free |
|---|---|---|---|---|---|---|
| first order | PI | cancellation | $T_1/\lambda$ | $1/\lambda$ | – | $\lambda > 0$ |
| 2nd real, $T_1/T_2 \geq 2$ | PI | cancellation | $T_1/(4T_2)$ | $1/(4T_2)$ | – | none |
| 2nd real, $T_1/T_2 \leq 2$ | PI | coincident, $\sigma = (T_1+T_2)/(3T_1T_2)$ | $(T_1^2-T_1T_2+T_2^2)/(3T_1T_2)$ | $(T_1+T_2)^3/(27T_1^2T_2^2)$ | – | none |
| 2nd real, any ratio | PID | cancellation | $(T_1+T_2)/\lambda$ | $1/\lambda$ | $T_1T_2/\lambda$ | $\lambda > 0$ |
| 2nd real, $T_1/T_2 < 2$ | PID | coincident | $3T_1T_2\sigma^2 - 1$ | $T_1T_2\sigma^3$ | $3T_1T_2\sigma - (T_1+T_2)$ | $\sigma_0 \leq \sigma \leq 1/T_1$ |

| Plant | C | Branch | $K_p$ | $K_i$ | $K_d$ | Free |
|---|---|---|---|---|---|---|
| 2nd complex, $\zeta > \sqrt{3}/2$ | PI | coincident, $\sigma = 2\zeta/(3T)$ | $(4\zeta^2-3)/3$ | $8\zeta^3/(27T)$ | – | none |
| 2nd complex | PID | coincident | $3T^2\sigma^2 - 1$ | $T^2\sigma^3$ | $3T^2\sigma - 2\zeta T$ | σ in window |
| 3rd real, separated | PID | cancellation | $(T_1+T_2)/(4T_3)$ | $1/(4T_3)$ | $T_1T_2/(4T_3)$ | none |
| 3rd real, clustered | PID | coincident, σ from (17) | $4T_1T_2T_3\sigma^3 - 1$ | $T_1T_2T_3\sigma^4$ | $6T_1T_2T_3\sigma^2 - (T_1+T_2+T_3)$ | none |
| 3rd, complex + real | PID | coincident, $\sigma = (T + 2\zeta T_3)/(4TT_3)$ | $4T_3T^2\sigma^3 - 1$ | $T_3T^2\sigma^4$ | $6T_3T^2\sigma^2 - (T_3+2\zeta T)$ | none |

Two constructions realize the binomial target; Table 2 records which one applies in each case, and Table 3 collects the resulting closed-form gains derived in Sections 3 and 4. In the **cancellation** construction the controller zeros sit on real plant poles and are removed from the loop, leaving a pure binomial closed loop with no setpoint zero; a PI cancels one pole and a PID cancels two. A free closed-loop pole survives only when the controller cancels *every* plant pole (a PI on a first-order plant or a PID on a second-order plant), so the lone remaining pole, together with the integrator, may be placed at will. When one real pole is left uncancelled (a PI on two real poles, or a PID on three), that pole and the integrator are pinned to a critically damped double pole and no free parameter remains. The **coincident** construction applies whenever the plant carries a complex pair, which a real-zeroed PI cannot cancel: all $m + 1$ closed-loop poles are gathered at a single $-\sigma$ and the controller zeros are retained. There σ is fixed by matching the $s^m$ coefficient, except that a PID on a second-order plant carries one extra gain that releases σ. Both constructions are monotonic, so the faster is selected: cancellation for well-separated real poles, coincident for a complex pair (and for tightly clustered real poles, where the retained zeros of the coincident design add lead and shorten settling). Table 4 lists one representative design per case; every response is strictly monotonic with $M_t = 1$, and the gains follow from the closed-form expressions.

*Table 4. Representative designs (verified monotonic, Mt = 1).*

| Plant | Method | Controller | Poles | Kp | Ki | Kd | Ms | PM (°) | $IAE_d$ | $T_s$ (s) |
|---|---|---|---|---|---|---|---|---|---|---|
| 1/(2s+1) | cancellation | PI | free | 1.00 | 0.500 | 0 | 1.000 | 90 | 2.00 | 7.82 |
| 1/[(2s+1)(s+1)] | cancellation | PI | 0.25, 0.50 | 0.500 | 0.250 | 0 | 1.155 | 76.3 | 4.00 | 11.67 |
| 1/[(2s+1)(s+1)] | cancellation | PID | free | 1.500 | 0.500 | 1.000 | 1.000 | 90.0 | 2.00 | 7.82 |
| 1/(s²+1.8s+1), ζ=0.9 | coincident | PI | 0.6 | 0.080 | 0.216 | 0 | 1.235 | 72.8 | 4.63 | 12.12 |
| 1/(s²+1.2s+1), ζ=0.6 | coincident | PID | free | 0.470 | 0.343 | 0.900 | 1.000 | 96.9 | 2.92 | 10.71 |
| 1/[(4s+1)(2s+1)(s+1)] | cancellation | PID | 0.125, 0.25, 0.5 | 1.500 | 0.250 | 2.000 | 1.155 | 76.3 | 4.00 | 11.67 |
| 1/[(s+1)(s²+1.8s+1)] | coincident | PID | 0.7 | 0.372 | 0.240 | 0.140 | 1.216 | 73.4 | 4.17 | 11.06 |

The last column of Table 4 reports the integrated absolute error $IAE_d$ for a unit step load disturbance entering at the plant input. Every disturbance response in Table 4 is non-sign-changing, so $IAE_d = 1/K_i$ exactly; the integrated-error identity $IE = 1/K_i$ is classical [5], and here it becomes an exact IAE because the responses do not change sign. Disturbance attenuation is thus read directly from the integral gain: the cancellation designs, whose gains are pinned, trade load-disturbance attenuation for the deepest monotonic setpoint pole, while the free-parameter designs improve $IAE_d$ as σ, and with it $K_i$, is increased. It should be emphasized that the minimum-settling property established here holds within the two binomial constructions of Table 2; no optimality is claimed over the set of all monotonic PID designs, and a PID with retained zeros can settle faster at the cost of robustness, as the following comparison quantifies.

### *6.1 Comparison with SIMC*

For the third-order plant of Table 4, the proposed cancellation PID is compared with SIMC [4] applied through its intended workflow: the half rule reduces the plant to the second-order-plus-delay model $\tau_1 = 4$, $\tau_2 = 2.5$, $\theta = 0.5$; the series-form SIMC PID with $K_c = \tau_1/[k(\tau_c + \theta)]$, $\tau_I = \min(\tau_1, 4(\tau_c + \theta))$, $\tau_D = \tau_2$ is converted to parallel form and applied to the full third-order plant. Here $\theta = 0.5$ is an approximation artifact of the neglected lag rather than physical dead time; the half rule is how SIMC reaches delay-free higher-order plants, so the comparison sits in the intersection of the two methods' scopes; for plants with appreciable physical dead time SIMC applies directly, while the present method requires the model-reduction route discussed in the conclusion. Table 5 reports three choices of the SIMC tuning constant: the default $\tau_c = \theta$, the smallest $\tau_c$ found by simulation to give a strictly monotonic response ($\tau_c = 2.0$), and $\tau_c = 3.19$, which matches the maximum sensitivity of the proposed design ($M_s = 2/\sqrt{3}$). Fig. 5 shows the closed-loop responses. At matched robustness the proposed design settles 12% faster, while SIMC attenuates the load disturbance 8% better; this is the price of cancelling the slow poles, which removes them from the setpoint response but leaves them in the disturbance path. SIMC can also be tuned to settle faster than the proposed design while remaining monotonic ($\tau_c = 2.0$), at the cost of a higher $M_s$; the location of that monotonicity boundary, however, must be found by simulation, whereas the proposed gains are closed-form and monotonic by construction. The comparison thus exposes the essential trade-off: an a priori monotonicity guarantee with fixed, universal robustness, at a known and modest cost in load-disturbance attenuation. As a classical zero-overshoot reference, the Chien–Hrones–Reswick setpoint rule for 0% overshoot [21] is also reported, applied through its own reaction-curve identification ($L = 1.56$, $T = 8.62$ from the tangent construction; like the half-rule $\theta$, this apparent dead time represents the folded higher-order lags rather than physical delay; $L$ exceeds $\theta = 0.5$ because the reaction curve folds everything beyond a single lag, whereas the half rule retains two): on the true plant it produces 9.3% overshoot at $M_s = 1.475$ and settles more slowly than the proposed design; the rule honors its specification only for plants close to its first-order-plus-delay model, whereas the proposed gains are monotonic by construction. The proposed design also requires the smallest peak control effort in Table 5.

*Table 5. Comparison with SIMC on the third-order plant 1/[(4s+1)(2s+1)(s+1)] (SIMC designed from the half-rule model; all controllers applied to the full plant; $u_{max}$ is the peak controller output for the setpoint step with derivative filter $T_f = 0.01$ on every design).*

| Design | $\tau_c$ | $K_p$ | $K_i$ | $K_d$ | OS (%) | $T_s$ (s) | $IAE_d$ | $M_s$ | $u_{max}$ |
|---|---|---|---|---|---|---|---|---|---|
| Proposed (cancellation) | – | 1.500 | 0.250 | 2.000 | 0 | 11.67 | 4.00 | 1.155 | 202 |
| SIMC, default | 0.5 | 6.500 | 1.000 | 10.00 | 13.8 | 7.31 | 1.00 | 1.460 | 1007 |
| SIMC, monotonicity boundary | 2.0 | 2.600 | 0.400 | 4.000 | 0 | 6.96 | 2.50 | 1.216 | 403 |
| SIMC, matched $M_s$ | 3.19 | 1.764 | 0.271 | 2.713 | 0 | 13.24 | 3.69 | 1.155 | 273 |
| CHR 0% overshoot, setpoint [21] | – | 3.323 | 0.386 | 2.586 | 9.3 | 15.59 | 2.59 | 1.475 | 262 |

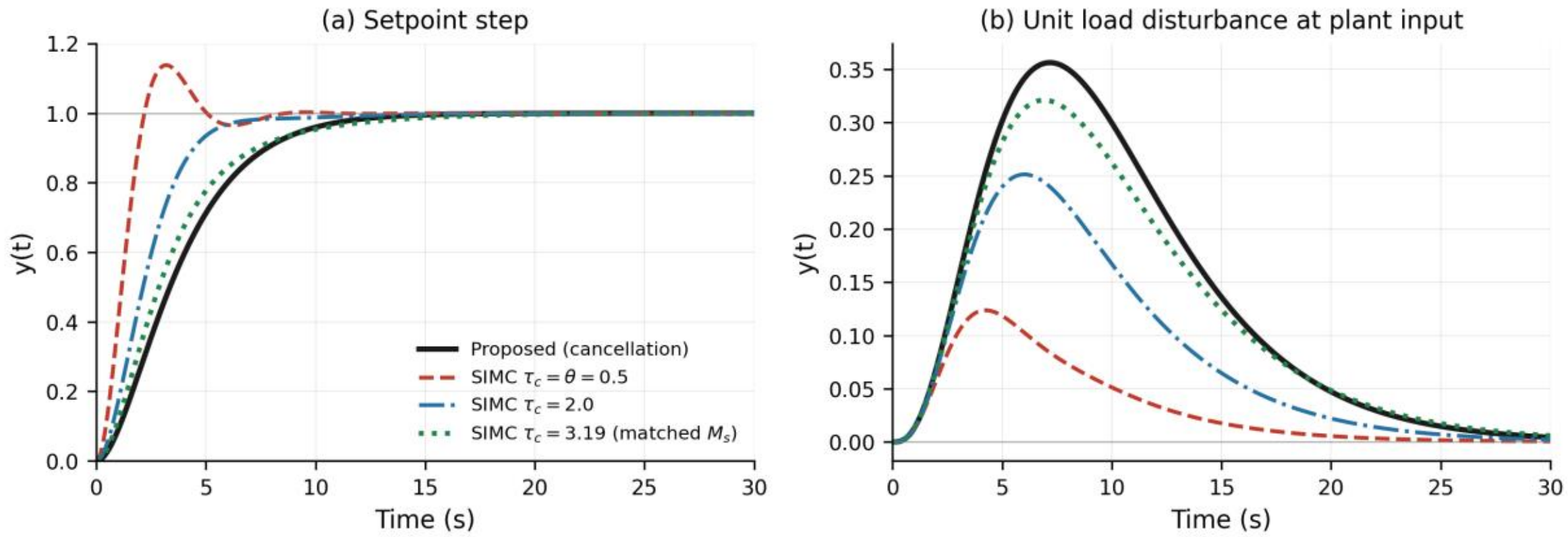

*Fig. 5. Third-order plant: proposed cancellation PID versus SIMC for three choices of $\tau_c$; (a) setpoint step, (b) unit load disturbance at the plant input.*

### 6.2 Comparison with explicit second-order formulas

The underdamped second-order plants of Table 4 permit a comparison with the explicit formulas of Albatran et al. [19], which fit PSO-optimized gains under a restricted deadbeat criterion. The two methods pursue different specifications: the deadbeat criterion of [19] explicitly forces a nonzero overshoot (0.1% ≤ PO < 2%), so a strictly monotonic response is excluded by construction, and no robustness constraint is imposed; the present designs are strictly monotonic with $M_t = 1$. Table 6 reports both, with [19] evaluated in its stated parallel form including the derivative filter $T_f = 1/(100\omega_n)$. For the PI case ($\zeta = 0.9$) the formulas of [19] settle faster (8.2 s versus 12.1 s) and attenuate the load disturbance better, at a slightly higher $M_s$ and without strict monotonicity; the proposed PI carries no free parameter, and its value lies in the one-line closed form and the guarantee. For the PID case ($\zeta = 0.6$) the gains of [19] are one to two orders of magnitude larger and produce a near-deadbeat response ($T_s = 0.07$ s with 2% overshoot and correspondingly large control effort), which the monotonic specification cannot match by definition. Within the strict specification, however, moving the free parameter to the top of its monotonic window, $\sigma = 1.151$, halves the settling time of the Table 4 design to 5.2 s and improves $IAE_d$ by a factor of 4.4 while keeping $M_s = 1$. The cost of the near-deadbeat speed of [19] appears in the control effort: with the same derivative filter the peak controller output is about 4600, against 90–230 for the proposed designs, a factor of 20 to 50, while the proposed PI never exceeds its steady-state output. Fig. 6 shows the responses.

*Table 6. Comparison with the explicit formulas of [19] on the underdamped second-order plants of Table 4 ($\omega_n = 1$; [19] in parallel form with derivative filter $T_f = 1/(100\omega_n)$; $u_{max}$ is the peak controller output for the setpoint step with the same filter on every PID).*

| ζ | Method | $K_p$ | $K_i$ | $K_d$ | OS (%) | $T_s$ (s) | $IAE_d$ | $M_s$ | $u_{max}$ |
|---|---|---|---|---|---|---|---|---|---|
| 0.9 | Proposed PI | 0.080 | 0.216 | 0 | 0 | 12.12 | 4.63 | 1.235 | 1.00 |
| 0.9 | Albatran et al. [19], PI | 1.194 | 0.598 | 0 | 0 | 8.17 | 1.67 | 1.294 | 1.34 |
| 0.6 | Proposed PID, σ = 0.7 | 0.470 | 0.343 | 0.900 | 0 | 10.71 | 2.92 | 1.000 | 90 |
| 0.6 | Proposed PID, σ = 1.151 | 2.974 | 1.525 | 2.253 | 0 | 5.21 | 0.656 | 1.000 | 228 |
| 0.6 | Albatran et al. [19], PID | 28.96 | 24.32 | 45.83 | 2.0 | 0.07 | 0.064 | 1.251 | 4612 |

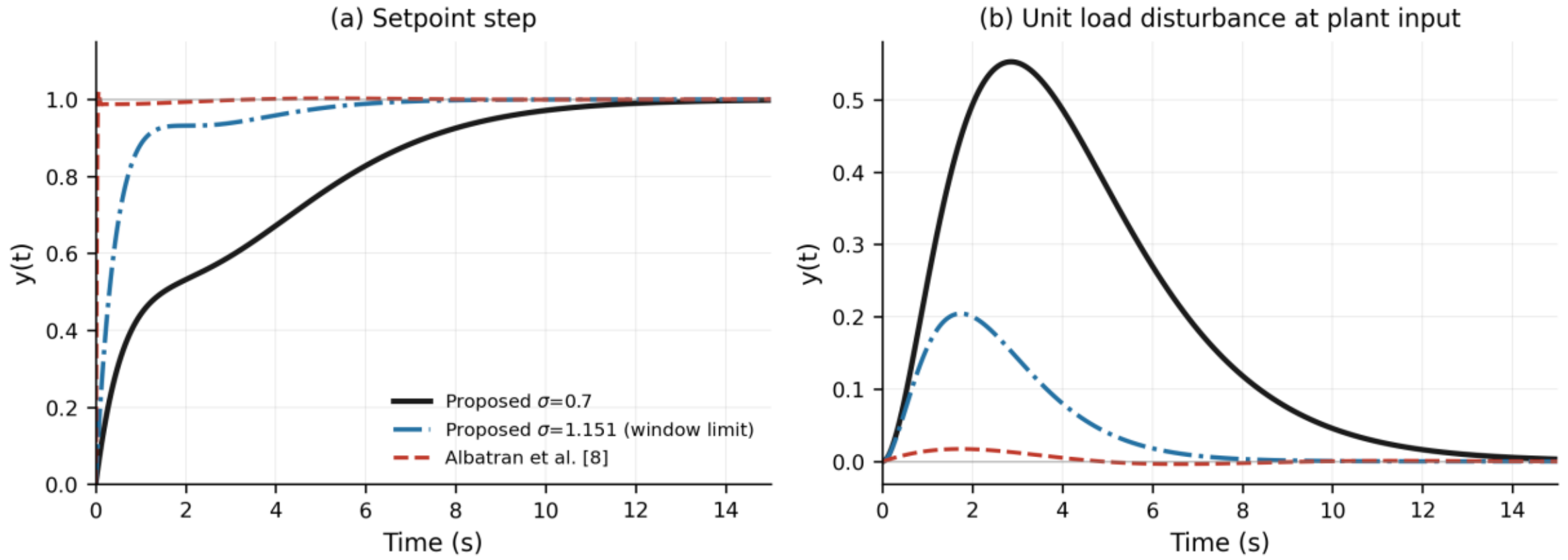

*Fig. 6. Underdamped second-order plant (ζ = 0.6): proposed PID at σ = 0.7 and at the window limit σ = 1.151, versus the formulas of [19]; (a) setpoint step, (b) unit load disturbance at the plant input.*

## 7. Conclusion

A closed-form PI/PID tuning has been given for all-pole plants up to third order that is strictly monotonic and minimum-settling, covering real and complex poles. The design targets the binomial closed loop; realizing it exactly forces the controller zeros to be cancelled, which requires the controller numerator to divide the plant denominator, while for clustered poles the coincident branch with retained zeros settles faster. The construction holds for any stable plant up to second order with PI and third order with PID. Monotonicity guarantees $M_t = 1$ and hence $M_s \leq 2$, PM ≥ 60° and GM ≥ 6 dB, with universal constants for the binomial family. Compared with SIMC at matched maximum sensitivity on the third-order example, the design settles faster at a modest cost in load-disturbance attenuation; the minimum-settling property holds within the two binomial constructions rather than over all monotonic PID designs.

For fourth order and above the residual factor acquires a complex pair of damping $\sqrt{3}/2$ that a generic plant does not contain, so the exact cancellation cannot be performed; monotonic PID designs still exist but carry zeros, settle more slowly, and form a region rather than a closed-form point. The practical route for such plants is model-order reduction: approximate the plant by a third-order (or lower) surrogate (for example by the half-rule, folding the fast dynamics into an effective lag) and apply the present closed-form tuning to the surrogate. This keeps the controller a standard PI/PID and preserves the monotonicity and robustness guarantees on the dominant response.